\documentclass[showpacs,preprintnumbers,preprint,aps,amsmath,amssymb]{revtex4}
\usepackage{epsfig}

\def\dmsqs{\Delta m^2_{21}}
\def\dmsqa{\Delta m^2_{32}}

\def\vcc{V_{cc}}
\def\ta{\theta_{12}}
\def\tb{\theta_{23}}
\def\tc{\theta_{13}}
\def\Dsol{\Delta_{21}}
\def\Datm{\Delta_{32}}
\def\eV{\rm{eV}}

\def\be{\begin{equation}}
\def\ee{\end{equation}}
\def\bea{\begin{eqnarray}}
\def\eea{\end{eqnarray}}
\def\bes{\begin{subequations}}
\def\ees{\end{subequations}}
\def\ba{\begin{array}}
\def\ea{\end{array}}
\def\beq{\begin{equation}}
\def\eeq{\end{equation}}
\def\barr{\begin{eqnarray}}
\def\earr{\end{eqnarray}}

\def\dmsq{\Delta m^2}

\def\lsim{\mathrel{\rlap{\lower4pt\hbox{\hskip1pt$\sim$}}
    \raise1pt\hbox{$<$}}}                
\def\gsim{\mathrel{\rlap{\lower4pt\hbox{\hskip1pt$\sim$}}
    \raise1pt\hbox{$>$}}}                

\newcommand{\ch}{{\chi^2}}

\begin{document}


\title[]{Constraints on flavor-dependent long range forces \\ from
solar neutrinos and KamLAND}

\author{Abhijit Bandyopadhyay$^1$\footnote{abhi@theory.tifr.res.in},
Amol Dighe$^1$\footnote{amol@tifr.res.in}, and
Anjan S. Joshipura$^{2}$\footnote{anjan@prl.res.in}}
\affiliation{
1. Tata Institute of Fundamental Research,
Homi Bhabha Road, Mumbai 400 005, India \\
2. Physical Research Laboratory, Ahmedabad 380 009, India
}

\begin{abstract}
Flavor-dependent long range (LR) leptonic forces,
like those mediated by the $L_e-L_\mu$ or $L_e -L_\tau$ gauge bosons, 
constitute a minimal extension of the 
standard model that preserves its renormalizability. 
We study the impact of such interactions on 
the solar neutrino oscillations 
when the interaction range $R_{LR}$ is much larger than the
Earth-Sun distance.
The LR potential can dominate over the standard 
charged current potential inside the Sun
in spite of strong constraints on 
the coupling $\alpha$ of the LR force coming from the 
atmospheric neutrino data and laboratory search for new forces.
We demonstrate that the solar and atmospheric neutrino
mass scales do not get trivially decoupled
even if $\theta_{13}$ is vanishingly small.
In addition, for $\alpha \gsim 10^{-52}$ and normal hierarchy, 
resonant enhancement of $\theta_{13}$ results in nontrivial
energy dependent effects on the $\nu_e$ survival probability.
We perform a complete three generation analysis, and
obtain constraints on $\alpha$  through a global fit to
the solar neutrino and KamLAND data.
We get the $3\sigma$ limits 
$\alpha_{e\mu} < 3.4 \times 10^{-53}$ and 
$\alpha_{e\tau} < 2.5 \times 10^{-53}$ 
when $R_{LR}$ is
much smaller than our distance from the galactic center.
With larger $R_{LR}$,
the collective LR potential due to all the electrons in the
galaxy becomes significant and the constraints on $\alpha$
become stronger by upto two orders of magnitude.
\end{abstract}


\maketitle

\section{Introduction}
\label{intro}

The standard electroweak model is now a well-established theory but it is
believed to be incomplete and one expects some physics
beyond the standard model (SM) to exist. Most extensions of SM
postulate new
physics at scales higher than the electroweak scales starting from TeV to
the grand unification or Planck scale. There however exists an interesting
possibility that new physics may exist at scales below the electroweak
scale. This may arise from the existence of exactly or nearly massless
gauge \cite{foot} or Higgs bosons \cite{axion,singletm,tripletm} which have
remained invisible because of their very
feeble couplings to the known matter. Various scenarios involving
new physics at low energy and their possible signatures 
\cite{hall,mv,masso,mj1,mj2} have been studied.

Gauged extension of the SM is one possible scenario with new  physics
below the electroweak scale. Such a possibility is strongly
constrained theoretically from the renormalizability, however  
there exist \cite{foot} three possible
$U(1)_X$ gauge extensions of the standard model which are 
anomaly free with minimal matter content. These correspond 
to $X=L_e-L_\mu,~L_e-L_\tau,~L_\mu-L_\tau$.
The extra gauge boson corresponding to $U(1)_X$ may not have been
discovered if it is  very heavy or if it is (nearly) massless but couples
to the matter very weakly. The former possibility is analyzed in
\cite{foot,vga}. The latter possibility, first suggested in \cite{mj1}, is
strongly constrained by the search for the long range (LR)  forces
\cite{lr1,lr2}.

Unlike the gravitational force, the $U(1)_X$ induced force couples only to
the electron (and neutrino) density inside a massive object. As a
consequence, the resulting acceleration  experienced by an object depends
on its leptonic content and mass. Such forces that violate the
equivalence principle are
strongly constrained. In case of the force with a range of $\sim $AU, the
most stringent bound comes from lunar ranging \cite{lr1,lr2} which
measures
the differential acceleration of the Earth and moon towards the Sun. If
$\alpha$ denotes the strength of the long range potential then
these experiments imply $\alpha < 3.4 \times 10^{-49}$ ($2\sigma$) for a range 
$\lambda\gsim 10^{13}$cm.

The flavor-dependent long range force \cite{masso}  induced for example 
by $L_e-L_{\mu,\tau}$ \cite{mj1,mj2}  can still
influence neutrino oscillation in spite of such strong constraints on
$\alpha$. This happens because (i) the $X$-charge of the electron flavor
is opposite to that of muon or tau flavor, so that these twox
flavors propagate differently in matter and (ii) the large number of
electrons
(e.g. inside the Sun) and the long range of interaction compensates for
the smallness of coupling and gives rise to a significant potential. For
example, the electrons inside the Sun generate a potential $V_{e\beta}$ at 
the
Earth surface given by \cite{mj1}
\be \label{vlr}
V_{e\beta}^\odot (R_{es})=
\alpha_{e\beta} \frac{N_e^\odot}{R_{es}}\approx 1.3 \times 10^{-11} \eV
\left(\frac{\alpha_{e\beta}}{10^{-50}}\right) \; ,
\ee
where $\alpha_{e\beta}=\frac{g_{e\beta}^2}{4 \pi}$ corresponds to the 
gauge coupling of
$L_e-L_{\beta}~(\beta=\mu,\tau)$ symmetry which we will sometimes collectively refer to 
as $\alpha$.  
Here $N_e^\odot \sim 10^{57}$ is the total number of
electrons inside
the Sun \cite{bahcall} and $R_{es}$ is the Sun-Earth distance 
$\approx 7.6\times 10^{26}$ GeV$^{-1}$. 
This is to be compared with the typical value of
$\dmsq/E \sim 10^{-12}$ eV for the atmospheric neutrinos.
It follows that 
$V_{e\beta}$ can induce significant
corrections to neutrino oscillations at the Earth 
even for $\alpha\sim 10^{-50}$. 

One can define a parameter 
\be\label{xi}
\xi \equiv \frac{2 E V_{e\beta}}{\Delta m^2} \ee
which measures the effect of the long range force in any given neutrino
oscillation experiment. 
The bound on $\alpha$ from \cite{lr1,lr2} implies that
$\xi < 750$ in atmospheric or a typical long base line experiment,
while $\xi < 35 $ for the typical parameters of the KamLAND experiment.
Relatively large
values of $\xi$ tend to suppress the atmospheric neutrino oscillations.
The observed oscillations can then be used to put a stronger constraint on
$\alpha$ which were analyzed in \cite{mj1}. One finds the improved 
90\% C. L. bound
\be \label{atmbound}
\alpha_{e\mu} < 5.5 \times 10^{-52} \; , \;
\alpha_{e\tau} < 6.4\times 10^{-52}
\ee
in case of the $L_{e}-L_{\mu,\tau}$ symmetry respectively.

With the improved bound on $\alpha$ given in (\ref{atmbound}), the
value of $\xi$ for KamLAND becomes rather small: $\xi < 0.06$. 
So one expects the KamLAND results to be influenced by
the LR interactions to a very small extent.
However, the potential $V_{e\beta}$ at the surface of
the Sun is $V_{e\beta}^\odot (r_\odot) \approx 
2.8 \times 10^{-9} (\alpha_{e\beta}/10^{-50})$ eV, 
which may be compared with the MSW contribution 
$V_{CC} \approx 6.0 \times 10^{-12}$ eV
at $r = 0.05 ~r_\odot$.
Therefore one expects the long range potential to change
or disturb the MSW LMA solution of the solar neutrino problem 
\cite{lma}.
Note that the effects on the solar and KamLAND experiments
are qualitatively different,
since KamLAND only probes the potential at the Earth
given in eq.~(\ref{vlr}) while the solar
neutrinos experience  a long range potential that varies 
with the distance from the center of the Sun.
It is thus important to do a combined analysis of these 
two experiments.

The aim of this paper is to
discuss new physical effects associated with this force and also make a
quantitative analysis of the combined solar and KamLAND data to obtain
a bound on $\alpha$. It turns out that the long range potential produces
physically interesting and
quantitatively significant effects which can be used to
constrain its strength.
The bound obtained on $\alpha$ is more stringent than that obtained 
\cite{mj1} from the atmospheric results alone by 
more than an order of magnitude.
If $R_{LR} \gsim R_{\rm gal}$ where $R_{\rm gal}$ is our
distance from the galactic center
$\sim$ 10 kpc, the constraints become even stronger
by upto two orders of magnitude.

The plan of the paper is as follows. 
In Sec.~\ref{formalism} we present our basic formalism  
where we describe the main features of the LR potential
inside and outside the Sun.
In Sec.~\ref{analytic},  
we present an analytic discussion of our results on 
neutrino masses, mixing angles and the resonances they undergo
in case of the $L_e-L_\mu$ symmetry. 
The corresponding  analysis for the $L_e-L_\tau$ is similar and 
the relevant analytic expressions are 
given in the appendix~\ref{le-ltau}.
Sec.~\ref{numerical} analyzes the KamLAND 
and the solar neutrino data numerically to obtain
bounds on $\alpha$ for $R_{LR} \ll R_{\rm gal}$.
The case $R_{LR} \gsim R_{\rm gal}$ is analyzed in Sec.~\ref{galaxy}.
A summary of the results is given in Sec.~\ref{summary}.
In addition, appendix~\ref{supernova} gives a brief discussion 
of the impact of the LR 
potential on the neutrinos from a core collapse supernova.

\section{Formalism}
\label{formalism}

We consider the standard electroweak model with its minimal fermionic
content but assume the presence of 
an additional gauged $U(1)_X$ symmetry.
The cancellation of anomalies requires
$X=L_e-L_\mu,L_e-L_\tau$ or
$L_\mu-L_\tau$ \cite{foot}. The last symmetry does not play any
significant role in the solar neutrino oscillations 
because of the absence of muons or tau leptons inside the Sun 
(or Earth). 
We will therefore concentrate on the first two and the couplings
of the mediating vector bosons. The value of $\alpha$ is 
positive in this case. 

The observed neutrino oscillations imply that the
$U(1)_X$ gauge symmetry cannot be an exact symmetry in nature. This is
easy to argue. If it were exact, then the effective five dimensional
neutrino mass operator following from any mechanism (e.g. seesaw)
would be invariant under it. Consider the case of $L_e-L_\mu$. 
Invariance under this dictates the following structure for the 
effective neutrino mass matrix:
\be
\label{meff}
m_{eff}=\left( \ba{ccc} 0&m_{e\mu}&0\\ m_{e\mu}&0&0\\
0&0&m_{\tau\tau} \ea \right) \; . \ee
This structure implies a Dirac and a Majorana neutrino which remain
unmixed and therefore cannot give any neutrino oscillations. Thus
$L_e-L_\mu$ needs to be broken. 
The symmetry breaking scale required to generate the solar scale
$\Delta m_{12}^2$ would be $\Delta m_{12}^2/m_{e\mu}$.
With $\Delta m_{12}^2\sim 10^{-4}$ eV$^2$ 
corresponding to the solar mass
difference and $m_{e\mu} \sim 0.1$ eV corresponding to 
the degenerate neutrino mass,
 $\Delta m_{12}^2/m_{e\mu}$ is required to be at least 
$10^{-3}$ eV.
A similar conclusion also holds in the case of the 
unbroken $L_e-L_\tau$ symmetry.

The size of the  $U(1)_X$ breaking  as required above can be
consistent with a nearly massless  gauge boson  
since the corresponding
coupling $g$ in this case is required to be very small ($\lsim 10^{-24}$)
from the search of the long range forces \cite{lr1,lr2}. The required
smallness of the coupling also ensures that the relatively large $U(1)_X$
breaking in the neutrino sector is consistent with a very light gauge
boson. In fact, a Higgs vacuum expectation value of a few GeV can lead to a
gauge boson corresponding to the Earth-Sun range with $g\sim 10^{-26}-10^{-27}$
and can imply a relatively large neutrino splitting \cite{mj1}.

\begin{figure}
\centerline{\psfig{figure=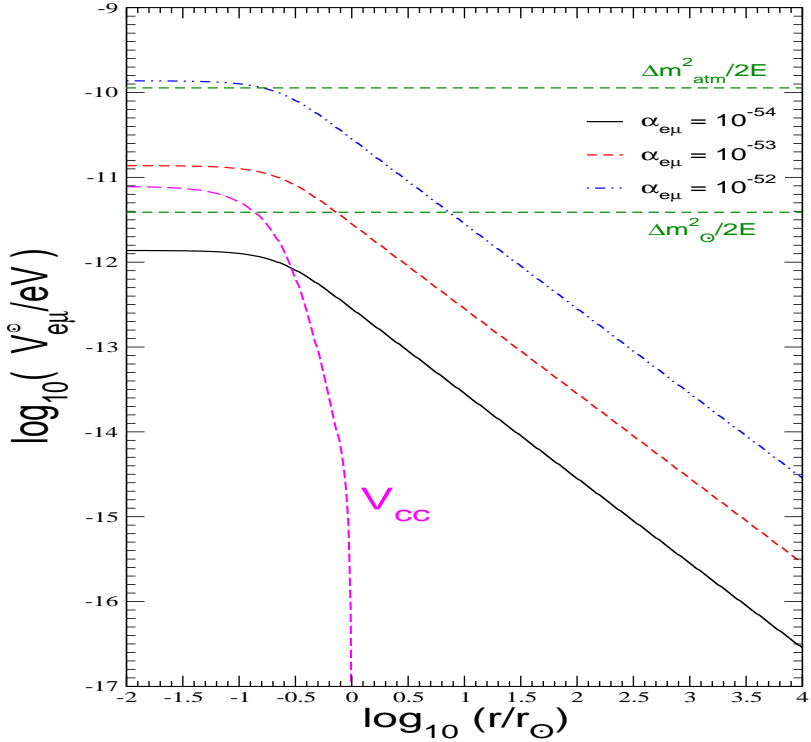,height=10cm,width=12cm,angle=0}}
\caption{Comparison of the MSW potential $V_{cc}$ 
and the LR potential $V_{e\mu}^\odot$ due to the solar electrons from the
solar core all the way to the Earth ($r/r_\odot \approx 215$) and beyond. 
The $(\dmsq/2E)$ values
corresponding to $E=10$ MeV are also shown.
\label{fig:vemu}
 }
\end{figure}

The most significant effect of the light  gauge bosons
would be in the solar neutrino oscillations. The coupling of the
solar electrons to the $L_e-L_{\mu,\tau}$ gauge bosons would
generate a long range potential. If $n_e(r)$ denotes the spherically
symmetric electron number density 
inside the Sun then the long range
potential is given by
\be
V_{e\beta}^\odot (r < r_\odot) = 4 \pi \alpha_{e\beta}
\int_r^\infty \frac{dr'}{r'^2} ~ \int_0^{r'} r''^2 n_e(r'') dr'' \; .
\label{vebeta<}
\ee
Outside the Sun,
\be \label{vebeta>}
V_{e\beta}^\odot (r > r_\odot)= \frac{4 \pi\alpha_{e\beta}}{r} 
\int_0^{r_\odot} r''^2 n_e(r'')dr'' =
\frac{\alpha_{e\beta}}{r} N_e^\odot \; .
\ee
The approximate profile 
\beq
n_e(r)\sim 245~ N_A~ 10^{-10.54 ~(r/r_\odot)} ~{\rm cm}^{-3}
\eeq
of the solar density \cite{bahcall} implies that $V_{e\beta}$
is a monotonically decreasing function, which is
inversely proportional to $r$ when outside the Sun.
This behavior is shown in
Fig.~\ref{fig:vemu} which is obtained using the actual electron density 
profile in the Sun.
It is seen that 
$V_{e\beta}^\odot$ dominates over the MSW potential $V_{CC}$
inside the Sun for $\alpha \gsim 10^{-53}$.
Moreover, it does not abruptly go to zero outside the Sun 
like $V_{CC}$, but decreases inversely with $r$, ultimately
reaching the value given in eq.~(\ref{vlr}) at the surface of the
Earth.
When $V_{e\beta} \gsim V_{CC}$ inside the Sun, 
the resonance is shifted outwards (sometimes
even outside the Sun) and its adiabaticity may be affected.

The contribution $\tilde{V}_{e\beta}^E$
of the electrons inside the Earth can be calculated in a similar 
fashion. Roughly, one finds that at the Earth surface,
\beq
\frac{{V}_{e\beta}^E}{V_{e\beta}^\odot}\approx
\frac{M_E}{M_\odot}\frac{R_{es}}{R_E}\approx 10^{-1} \; ,
\eeq
where $M_{\odot}(M_E)$ and $R_{es} (R_E)$ respectively refer to the
mass of the Sun (Earth) and the Earth-Sun distance 
(radius of the Earth).
Thus the solar long range potential dominates over the terrestrial
contribution and we will neglect the latter.
As long as $R_{LR} \ll R_{\rm gal}$, this is the dominant potential
affecting the propagation of solar neutrinos.

When $R_{LR} \gsim R_{\rm gal}$, the collective potential due to
all the electrons in the galaxy may become significant. 
The mass of the Milky Way is (0.6 -- 3.0) $\times 10^{12}$ solar masses,
which is mostly concentrated in the center of the galaxy.
The baryonic contribution to the galactic mass may be
estimated to be ${\cal O}(10\%)$. 
The center of the galaxy is $\sim$ 10 kpc away from the Sun.
We denote the galactic contribution to the potential $V_{e\beta}$ as
\beq
V_{e\beta}^{\rm gal} = b ~ \alpha_{e\beta}~ 
\frac{N^0_{e, {\rm gal}}}{R^0_{\rm gal}} \; ,
\label{b-def}
\eeq
where $N^0_{e, {\rm gal}}$ is taken to be $10^{12} N_e^\odot$
and $R^0_{\rm gal}$ to be 10 kpc.
The net LR potential is $V_{e\beta} = V_{e\beta}^\odot +
V_{e\beta}^{\rm gal}$.
The parameter $b$ takes care of our ignorance about the distribution of
the baryonic mass in our galaxy. With $R_{LR} \gsim R_{\rm gal}$,
we expect $0.05 < b \lsim 1$. The value of $b$ may be smaller 
if $R_{LR}$ is smaller. Clearly, $b=0$ would have the same effect
as $R_{LR} \ll R_{\rm gal}$.
With $b \neq 0$, the constraints on $\alpha$ become stronger,
as will be demonstrated in Sec.~\ref{galaxy}.

The screening length due to the antineutrinos present in the 
cosmic neutrino background is a few hundred kpc for $m_\nu \sim 0.1$
eV \cite{screening}. Therefore, for the galactic scale, the screening
plays no significant role. Over the Sun-Earth distance, even 
the possible local screening effects would be too small to have any 
effect \cite{mj1}.

In addition to the altered resonance structure inside and outside
the Sun, the mixing angles at the
Earth also differ from the corresponding vacuum values, 
with the result that 
both the solar and the KamLAND neutrinos get affected by the LR
potential.
An important point to note is that
this potential gives unequal contributions to two flavors 
($e$ and $\mu$ or $\tau$) simultaneously  unlike
in case of the charged current which contributes only to the electrons.
The third flavor gets no contribution.
As a consequence of this, the inclusion of three generations in the
solar analysis becomes necessary.

The appropriate Hamiltonian in the flavor basis describing the neutrino
propagation can be written as
\beq \label{hf}
H_f = R_{23}(\tb) R_{13}(\tc) R_{12}(\ta) H_0 
R_{12}^T(\ta) R_{13}^T(\tc)R_{23}^T(\tb) + V \; ,
\eeq
where $H_0$ refers to the effective Hamiltonian in the mass basis,
and $R_{ij}$'s are the rotation matrices in the $i$-$j$ plane.
Since the absolute masses of neutrinos play no part in the oscillation
phenomena, we can take the neutrino mass eigenvalues in vacuum to be
$0, \sqrt{\dmsqs}, \sqrt{\dmsqa}$ respectively, leading to
\beq
H_0 = {\rm Diag}(0, \Delta_{21}, \Delta_{32}) \; ,
\eeq
where $\Dsol \equiv \dmsqs/(2E)$ and $\Datm \equiv \dmsqa/(2E)$.
The rotation angles $\theta_{23}$ and $\theta_{12}$
are the vacuum mixing angles describing the atmospheric and the solar
neutrino oscillations respectively,
whereas $\theta_{13}$ is the third ``Chooz'' mixing angle.
We have assumed that no CP violation enters into picture here.

The matrix $V$ in (\ref{hf}) 
describes the combined contribution of the
charge weak currents as well as the long range forces. Explicitly,
\beq \label{v}
V = {\rm Diag}(V_{cc}+V_{e\mu}, -V_{e\mu}, 0) \; .
\eeq
The neutrino propagation is described by eq.~(\ref{hf}).
The corresponding antineutrino propagation is obtained by
the replacement $V\rightarrow -V$.

\section{Masses, Mixings and Resonances of Solar Neutrinos}
\label{analytic}

In order to analyze the propagation of solar neutrinos,
we rewrite eq.~(\ref{hf}) explicitly as
\be \label{mnuf} H_f=\Datm \left(\ba{ccc}
x s_{12}^2  +y_c+y_{e\mu}&
x c_{12}s_{12}c_{23} + s_{13} s_{23} &
-x c_{12}s_{12} s_{23} - s_{13} c_{23} \\
x c_{12}s_{12}c_{23} + s_{13} s_{23} &s_{23}^2+x c_{12}^2 
c_{23}^2-y_{e\mu}&c_{23}s_{23}(1-x c_{12}^2)\\
-x c_{12}s_{12}s_{23} - s_{13} c_{23}
&c_{23}s_{23}(1-x c_{12}^2)&c_{23}^2+x c_{12}^2 
s_{23}^2\\
\ea\right) \; ,\ee
where
\be \label{defy}
 x\equiv \frac{\Dsol}{\Datm} \approx 0.03 ~~,~~
y_{c}\equiv\frac{V_{cc}}{\Datm}=\frac{ 2 E
V_{cc}}{\Delta m_{32}^2}~~,~~
y_{e\mu}\equiv \frac{V_{e\mu}}{\Datm}=\frac{ 2 E
V_{e\mu}}{\Delta m_{32}^2}\; , \ee
and $s_{ij}\equiv \sin\theta_{ij}, c_{ij} \equiv \cos\theta_{ij}$.
Since $\theta_{13}$ is small ($\theta_{13} < 0.2$ \cite{lma}),
we have kept terms to only linear order in $s_{13}$.

Eq.(\ref{mnuf}) can be diagonalized through the unitary matrix 
\beq
U_m\equiv R_{23}(\theta_{23m}) R_{13}(\theta_{13m})R_{12}(\theta_{12m})
\eeq
such that
\be \label{dia} U_m^T H_f U_m= \frac{1}{2E} {\rm
Diag}(m_{1m}^2,m_{2m}^2,m_{3m}^2) \; . \ee 

The smallness of $x$ and $s_{13}$ can be used to approximately determine 
the matter dependent mixing angles of $U_m$ to the leading 
order in these parameters.
The angle $\theta_{23m}$ follows from 
the lower right $2\times 2$ block in (\ref{mnuf}):
\be \label{23m} \tan 2 \theta_{23m} \approx \frac{\sin 2\theta_{23}(1-x
c_{12}^2) }{\cos 2\theta_{23}(1-x c_{12}^2)+y_{e\mu}}\; . \ee
The subsequent diagonalization leads to
\beq
\tan 2 \theta_{13m}  \approx  
\frac{ 2 ~(x s_{12} c_{12} S + s_{13} C)}{C^2+x(c_{12}^2 S^2- s_{12}^2) 
-y_c-y_{e\mu}(1+\sin^2 \theta_{23m})}\; ,
\label{13m}
\eeq
where $S\equiv\sin(\theta_{23m}-\theta_{23})$ and
$C\equiv\cos(\theta_{23m}-\theta_{23})$.

As long as the denominator in eq.~(\ref{13m}) does not
vanish (which happens only in a very narrow range of $y_{e\mu}$
near $y_{e\mu}\approx 2/3$), we can take 
$\theta_{13m} \sim {\cal O}(x,s_{13})$. Neglecting terms 
that are quadratic or higher order in ${\cal O}(x,s_{13})$,
the effective Hamiltonian
in the new basis (after the 2-3 and 1-3 rotation) becomes
\beq 
\label{mfpp} H_{f''} \approx \Datm \left(\ba{ccc}
x s_{12}^2  +y_c+y_{e\mu} & x c_{12}s_{12} C -s_{13} S & 0 \\
x c_{12}s_{12} C -s_{13} S
& S^2 + x c_{12}^2 C^2 - y_{e\mu} c_{23m}^2 
& 0 \\
0 & 0 & C^2 + x c_{12}^2 S^2 - y_{e\mu} s_{23m}^2 \\
\ea\right) \; ,\ee
so that a 1-2 rotation through an angle $\theta_{12m}$,
given by
\beq
\label{12m} 
\tan 2 \theta_{12m} \approx  
\frac{2 ~(x s_{12} c_{12} C - s_{13} S)}{S^2 + x (c_{12}^2 C^2 - s_{12}^2) 
-y_c-y_{e\mu}(1+\cos^2 \theta_{23m})} \; ,
\eeq
completes the diagonalization.
The neutrino masses  are given as
\barr
m_{1m}^2 & \approx & 
\Delta_{32} E \left[
x(c_{12}^2 C^2 + S^2) + y_c
+y_{e\mu}\sin^2\theta_{23m}+ S^2 - D^{1/2}\right]~, 
\nonumber \\
m_{2m}^2 & \approx & 
\Delta_{32} E \left[
x(c_{12}^2 C^2 + S^2) + y_c
+y_{e\mu}\sin^2\theta_{23m}+ S^2 + D^{1/2}\right]~, 
\nonumber \\
m_{3m}^2 & = & 2 \Datm E (C^2 + x c_{12}^2 S^2 
- y_{e\mu} \sin^2\theta_{23m}) \; ,
\label{masses}
\earr
where
\beq
D=\left[S^2 + x (c_{12}^2 C^2 - s_{12}^2) 
-y_c-y_{e\mu}(1+\cos^2 \theta_{23m})\right]^2
+ 4 ~(x s_{12} c_{12} C - s_{13} S)^2\; .
\label{d-def}
\eeq

\begin{figure}
\centerline{\psfig{figure=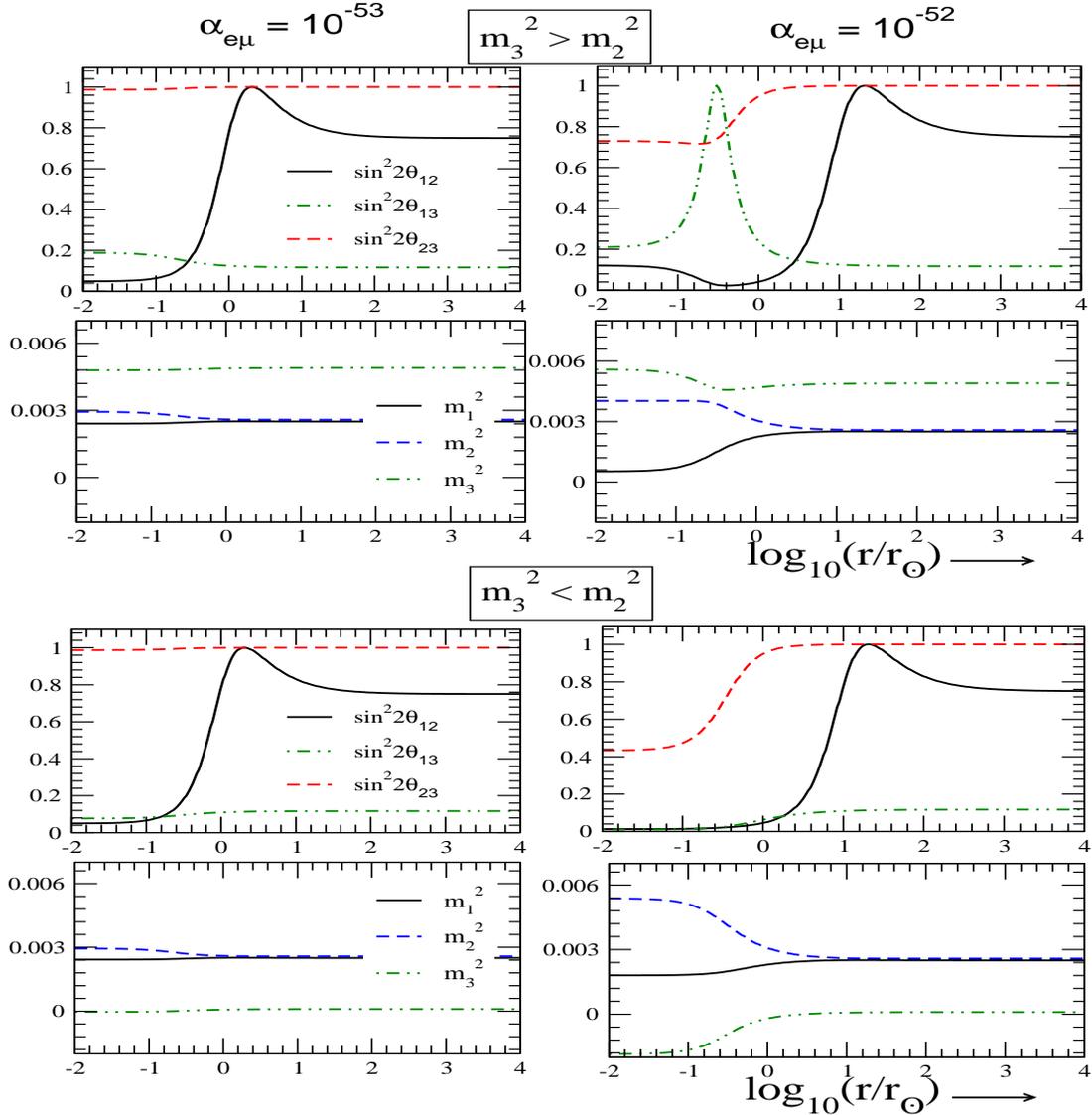,height=15cm,width=15cm,angle=0}}
\caption{The angles and $m_i^2$ values in matter for solar 
neutrinos 
for $E=10$ MeV, 
in the case $R_{LR} \ll R_{\rm gal}$. The $m_i^2$
values are correct up to an additive constant, so that only
their relative values have a physical significance.
\label{thetas}
}
\end{figure}

The above analytical results can be verified by the 
exact numerical results in Fig.~\ref{thetas}
where we show the angles and $m_i^2$ values in matter 
for different values of $\alpha$ for normal as well as
inverted hierarchy.

In this and the next section, we analyze the case $R_{LR} \ll
R_{\rm gal}$, so that the potential $V_{e\mu}$ is
as shown in Fig.~\ref{fig:vemu}.
As is apparent from the figure, 
the maximum value of $y_{e\mu}$ is given by
\be \label{yemumax}
\frac{ (y_{e\mu})_{max}}{\alpha} \approx 1.2
\times 10^{52}\left(\frac{E}{10 ~{\rm MeV}}\right) \ee 
for the best fit values of the atmospheric parameters.
Thus, at $E = 10$ MeV, we have $y_{e\mu} \approx 0.1$ for 
$\alpha = 10^{-53}$.
The left column in Fig.~\ref{thetas}
then corresponds to the range of $\alpha$ where
$(y_{e\mu})_{max} \ll 1$, 
and the right column corresponds to 
$(y_{e\mu})_{max} > 1$.

The propagation of solar neutrinos is qualitatively and quantitatively 
different depending on whether $(y_{e\mu})_{max}$ is large enough 
to cause resonant enhancement of $\theta_{13m}$ in (\ref{13m}).
The resonance occurs when $\alpha \gsim 10^{-52}$.
We therefore consider the two cases $\alpha \lsim 10^{-52}$ and
$\alpha \gsim 10^{-52}$ separately in the next two subsections.

\subsection{For $\alpha \lsim 10^{-52}$ }
\label{smallalpha}

For $\alpha \ll 10^{-52}$, we have $y_{e\mu} \ll 1$ 
and the atmospheric mixing angle gets only a 
small correction from the matter effects. Writing
$\theta_{23m}=\theta_{23}+\delta\theta_{23}$, we see that
$\delta \theta_{23}\approx -y_{e\mu} \sin 2\theta_{23}/2 $
and $\theta_{23m}$ remains 
close to its vacuum value, $\theta_{23m} \approx \pi/4$. 
With a higher $\alpha$ value, the deviation $\delta \theta_{23}$ 
becomes appreciable and results in the reduction of 
$\sin^2 2\theta_{23m}$ as shown in Fig.~\ref{thetas}. 

The angle $\theta_{13m}$ has contributions from two sources: 
from finite $\theta_{13}$ in vacuum as well as the additional
contribution from the term $x s_{12} c_{12} S$.
The latter is doubly suppressed because of the smallness of 
$x$ as well as $S \approx \delta\theta_{23}$, and can be
neglected as long as $s_{13} > x S$ .
One may then take $\theta_{13m} \sim {\cal O}(s_{13})$
since the resonant enhancement of $\theta_{13m}$
anyway does not occur for $\alpha < 10^{-52}$.

In the limit $\theta_{13m} \to 0$, the third mass eigenstate 
decouples and the scenario reduces to 2$\nu$ mixing,
as can be seen from eq.~(\ref{mfpp}).
However, note that the effective matter potential is 
\beq
{\cal V}_{12} \approx V_{cc} + V_{e\mu} (1+\cos^2 \theta_{23m}) \; , 
\label{v12}
\eeq
and not $V_{cc}+ 2 V_{e\mu}$ as would have been taken in a naive
2-generation analysis. Thus, the effect of the third neutrino and
its mixing is inescapable here. However, it only appears through
the factor $(1+c_{23m}^2)$ in eq.~(\ref{v12}), and the mass of the
third neutrino or the mass hierarchy is immaterial for the effective
2$\nu$ analysis.
This may be verified from the left column of fig.~\ref{thetas}.

The most important effect of the LR potential is for the
solar angle. Eq.~(\ref{12m}) gives the resonance condition
\be \label{solarreso} \Delta m_{21}^2 \cos 2 
\theta_{12}\approx 2E \left[V_{cc}+V_{e\mu}
(1+\cos^2 \theta_{23m})\right] \; ,\ee 
which differs from the MSW condition by an addition of 
the term involving $V_{e\mu}$.
For $\alpha \gsim 10^{-53}$, the $V_{e\mu}$ 
contribution dominates over $V_{cc}$ and
changes the MSW resonance picture significantly.
The resonance is shifted away from the center 
as $\alpha$ increases. Eventually for some value of 
$\alpha$ the resonance gets shifted outside the Sun where its 
behavior is solely determined by the LR potential. 
For $\alpha > 10^{-53}$, neutrinos with $E=10$ MeV 
encounter resonance outside the Sun in case of  
the best fit values of the neutrino mass parameters 
obtained in the standard analysis \cite{lma}.

Addition of the $V_{e\mu} (1+\cos^2 \theta_{23m})$ term 
to $V_{cc}$ makes the variation of the total 
potential smoother than the normal MSW potential with the result that 
the transition becomes more adiabatic than the corresponding case 
without 
the LR. In particular, when the resonance occurs outside the Sun then
$V_{e\mu}^\odot \sim \frac{\alpha N_e^\odot}{r}$ 
and the adiabaticity parameter 
at the resonance is given by
\bea 
\gamma_{L}&\equiv& \frac{\Delta m_{12}^2}{2 E} 
\frac{\sin^2 2\theta_{12}}{\cos 2\theta_{12}}
\left| \frac{1}{{\cal V}_{12}} \frac{d {\cal V}_{12}}{dr} 
\right|_{res}^{-1}
\nonumber \\
&\approx & \alpha_{e\mu} N_e^\odot 
\tan^2 2 \theta_{12} (1+\cos^2 \theta_{23})~\approx 1.4 \times
10^{58}~ \alpha_{e\mu} \; .
\label{gammaL}
\eea
The value of $\gamma_{L}$ is independent of the neutrino masses, 
energy and position of 
the resonance and is solely determined by $\alpha$ and the vacuum mixing 
angles. For the standard values of the latter, the resonance is found to 
be highly  adiabatic: $\gamma_{L} \gg 1$ for $\alpha > 10^{-57}$.  
In general, if $P_L (E) \equiv exp(-\pi \gamma_L/2)$
is the probability that $\nu_{1m}$ and $\nu_{2m}$ convert to
each other while passing through the resonance,  
the net survival probability of $\nu_e$ is
\barr
P_{ee}(E) & = &
(1-P_L) \cos^2 \theta_{13P} \cos^2\theta_{12P} 
\cos^2 \theta_{13E} \cos^2\theta_{12E} 
 \nonumber \\
& + & P_L \cos^2 \theta_{13P} \sin^2\theta_{12P} 
\cos^2 \theta_{13E} \cos^2\theta_{12E} 
 \nonumber \\
& + & (1 - P_L) \cos^2 \theta_{13P} \sin^2 \theta_{12P} 
\cos^2\theta_{13E} \sin^2\theta_{12E} 
\nonumber \\
& + & P_L \cos^2 \theta_{13P} \cos^2 \theta_{12P} 
\cos^2\theta_{13E} \sin^2\theta_{12E} 
\nonumber \\
& + & \sin^2 \theta_{13P} \sin^2\theta_{13E} \; .
\label{pee}
\earr
Here $\theta_{ijP}$ and $\theta_{ijE}$ 
are the values of $\theta_{ijm}$ at
the neutrino production point and at the Earth respectively. 
The energy dependence of $P_L$ as well as all the angles is implicit.
Note that since $\theta_{13P}, \theta_{13E} \sim {\cal O}
(\theta_{13})$, the last term may be neglected if we neglect
terms of ${\cal O}(\theta_{13}^4)$ or smaller.

\subsection{For $\alpha {\gsim} 10^{-52}$}
\label{largealpha}

For $\alpha \sim 10^{-52}$, the value of $y_{e\mu}$ is large enough
so that $\sin^2 2\theta_{23m}$ gets unacceptably suppressed
through eq.~(\ref{23m}).
This also suppresses the atmospheric neutrino flux and results in the
bounds on $\alpha$ discussed in \cite{mj1}.

For solar neutrinos, the $\nu_{1m}$-$\nu_{2m}$ resonance
as described in the previous section occurs, but in addition 
the angle $\theta_{13m}$ gets resonantly enhanced when
\beq 
C^2 + x (S^2 c_{12}^2 - s_{12}^2 ) -y_c 
- y_{e\mu} (1+ \sin^2 \theta_{23m}) \approx 0 \; .
\label{13-res}
\eeq
This happens when $y_{e\mu} \approx 2/3$. The sign of $y_{e\mu}$
also needs to be positive, so the resonance occurs only for
normal hierarchy. 
For the inverted hierarchy, there is no resonance for $\nu_e$
and eq.~(\ref{pee}) gives the correct expression for their
survival probability.

In the resonance region, the effective Hamiltonian matrix
(\ref{hf}) can no longer be diagonalized through the simple
procedure described in the beginning of Sec.~\ref{analytic},
and the mixing angles have to be computed numerically.
However, this happens only in a small range of $y_{e\mu}$
around $y_{e\mu}\approx 2/3$: the width of the resonance
region may be estimated to be 
$\delta y_{e\mu} \approx (2/3) s_{13}$. The expressions
(\ref{12m})-(\ref{d-def}) are valid everywhere outside this region.

The $\theta_{13m}$ enhancement corresponds to the 
$\nu_{2m}$-$\nu_{3m}$ level crossing, with an effective
potential
\beq
{\cal V}_{23} = V_{cc} + V_{e\mu} (1+\sin^2 \theta_{23m}) \; .
\label{v23}
\eeq
When the hierarchy is normal, only a fraction of the $\nu_e$ 
that are produced
mainly as $\nu_{2m}$ inside the Sun survive the
$\nu_{2m}$-$\nu_{3m}$ resonance.
The adiabaticity at this resonance, 
which strongly depends on $\theta_{13}$, affects the net survival
probability of $\nu_e$:
\barr
P_{ee}(E) & = &
\cos^2 \theta_{13P} \cos^2\theta_{12P} 
\cos^2 \theta_{13E} \cos^2\theta_{12E} 
 \nonumber \\
& + & (1 - P_H) \cos^2 \theta_{13P} \sin^2 \theta_{12P} 
\cos^2\theta_{13E} \sin^2\theta_{12E} 
\nonumber \\
& + & P_H \sin^2 \theta_{13P} 
\cos^2 \theta_{13E} \sin^2\theta_{12E} 
\nonumber \\
& + & (1 - P_H) \sin^2 \theta_{13P} \sin^2\theta_{13E} 
\nonumber \\
& + & P_H \cos^2 \theta_{13P} \sin^2 \theta_{12P}
\sin^2 \theta_{13E} \; ,
\label{pee-3nu}
\earr
where $P_H(E)$ is the probability that $\nu_{2m}$ converts to $\nu_{3m}$
after traversing through this resonance. 
\begin{figure}
\psfig{file=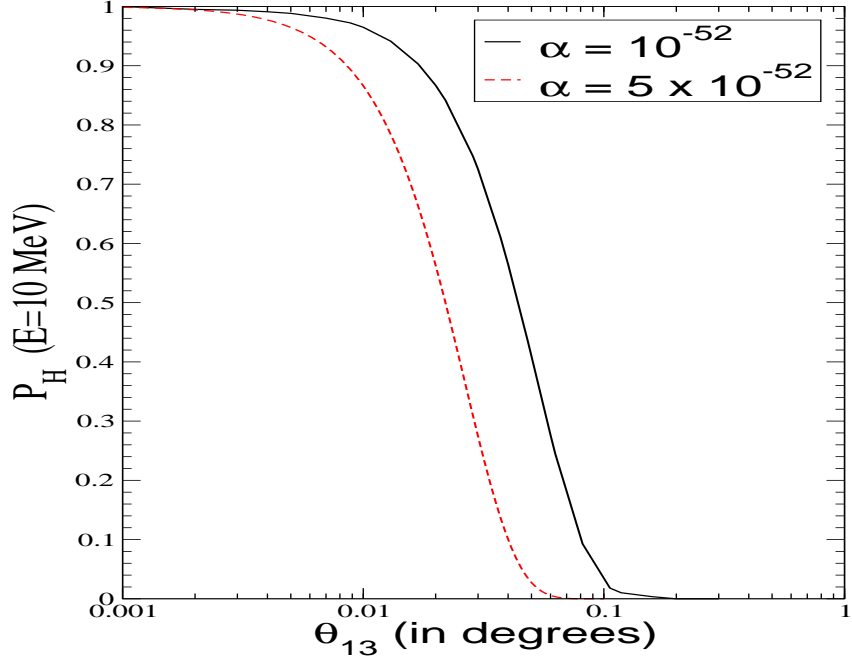,height=10cm,width=12cm}
\caption{  $\theta_{13}$ dependence of $P_H$
for various values of $\alpha$ for $E=10$ MeV in the case 
$R_{LR} \ll R_{\rm gal}$.
\label{p-t13}}
\end{figure}
Here we have used the earlier result that for $\alpha \gsim 10^{-52}$
the $\nu_{1m}-\nu_{2m}$ resonance is outside the Sun and is
always adiabatic [see eq.~(\ref{gammaL})].
The energy dependence of $P_H$ as well as all the angles is implicit.

The value of $P_H$ is given by
\beq
P_H \approx \exp \left[ - \frac{\pi}{2} 
\left| \frac{m_3^2 - m_2^2}{2 E~ d\theta_{13m}/dr} \right|_{\rm res}
\right] \; .
\label{p-def}
\eeq
Clearly, if $P_H \approx 0$, 
the expression (\ref{pee-3nu}) reduces to 
(\ref{pee}), and the results of the 2$\nu$ analysis 
stay valid.
In general $P_H \approx 0$ at high values of $\theta_{13}$.
In Fig.~\ref{p-t13}, we show the $\theta_{13}$ dependence of $P_H$
for various values of $\alpha$ for $E=10$ MeV. 
At $\alpha = 10^{-52}$, the value of $P_H > 0.1$ 
for $\theta_{13} < 0.08^\circ$, which is when the 
survival probability is affected significantly.
For larger $\alpha$, the value of $P_H$ becomes significant
for lower $\theta_{13}$ values.
In the range where $0.1 < P_H < 0.9$ (the semi-adiabatic range), 
$P_H$ is also highly energy 
dependent, as can be seen from (\ref{p-def}).

The analytic discussion above reveals that the LR potential makes 
important contribution to the solar neutrino problem and a detailed 
numerical analysis is needed to obtain constraints on this potential.
We turn to this analysis in the next section.

\section{Constraints from solar neutrinos and KamLAND}
\label{numerical}

To find the best fit values of the oscillation parameters and
$\alpha$ from a statistical analysis of the experimental
data, we employ the $\ch$ minimization technique
with covariance approach for the errors.
For analysis of the total event rate data from all the experiments,
 the $\ch$ function is defined as
\begin{eqnarray}
\ch_{\rm rates} &=& \sum_{i,j=1}^{N_{\rm expt}}
(P^{\rm th}_i - P^{\rm expt}_i) \left[ (\sigma_{ij}{}^{rates})^2
\right]^{-1}
(P^{\rm th}_j - P^{\rm expt}_j) \; ,
\label{chrate}
\end{eqnarray}
where P$_i^\xi$ ($\xi$ = th or expt) denotes the total event rate for
the $i^{\rm th}$ experiment. Both the theoretical and experimental values
of the fitted quantities are normalized relative to 
the standard solar model (SSM) predictions.
The error matrix $(\sigma_{ij}{}^{rates})^2$ contains the experimental
and theoretical uncertainties along with their correlations. 
Theoretical uncertainties include the uncertainties in the capture
cross sections, which are uncorrelated between different experiments
and the astrophysical uncertainties from the SSM predictions which are
correlated between different experiments. The correlations are being
evaluated using the procedure of \cite{ana2g_lisicorr}.

For the analysis of
any spectral data (recoil energy spectra or zenith angle
spectra), the $\ch$ is defined as
\begin{eqnarray}
\ch_{\rm spec} &=&
\sum_{i,j=1}^{N_{\rm bins}}
(S^{\rm th}_i - S^{\rm expt}_i) \left[ (\sigma_{ij}{}^{spec})^2
\right]^{-1}
(S^{\rm th}_j - S^{\rm expt}_j) \; ,
\label{chspec}
\end{eqnarray}
where S$_i^\xi$ ($\xi$=th or expt) is the number of events in the
$i^{\rm th}$ bin of the spectrum. The error matrix $(\sigma_{ij}{}^{spec})^2$
for the spectral data includes the statistical error, correlated and
uncorrelated systematic errors in the different bins and the error due
to the calculation of the neutrino energy spectrum from SSM.
                                                                                
For a global analysis of the solar data -- rates from Cl, Ga
experiment, spectral data from SuperKamiokande (SK) and 
Sudbury Neutrino Observatory (SNO): both D$_2$O and salt phase,
and KamLAND data --
the relevant $\ch$ is given by
\begin{eqnarray}
\ch = \ch_{\rm Cl,Ga~rates} + \ch_{\rm SK~spec} + \ch_{\rm SNO~spec}
+ \ch_{\rm KamLAND} \; .
\label{chisqare}
\end{eqnarray}

Note that only solar neutrino observations would not have been 
able to put strong constraints on $\alpha$: as long as 
there is no $\nu_{2m}$-$\nu_{3m}$
level crossing and the $\nu_{1m}$-$\nu_{2m}$ resonance is
adiabatic,  the net $\nu_e$ survival probability 
(\ref{pee}) is a function of $\Delta m_{12}^2/\alpha$ 
for a large $\alpha$, i.e. for $V_{cc}<< V_{e\mu}$. 
As a result, one can fit the solar data by increasing 
the value of $\Delta m_{12}^2$ when $\alpha$ is increased 
and a strong bound on $\alpha$ would not follow.

However, the data from KamLAND restricts $\Delta m^2_{21}$
to a very small range and plays a crucial role in 
constraining $\alpha$. 
We use eqs.~(\ref{pee}) and (\ref{pee-3nu}) for the survival 
probability of solar neutrinos. 
The $\bar{\nu_e}$ survival probability
in KamLAND is given by
\barr
P_{\bar{e}\bar{e}}^{KL}
&=&
1 -
\cos^4\theta_{13}
\left[
\sin^22\theta_{12}\sin^2\left(\frac{\Delta m^2_{21} L}{4E}\right)
\right]
-\sin^22\theta_{13}\sin^2\left(\frac{\Delta m^2_{31} L}{4E}\right)
\nonumber \\
&&
+ \sin^22\theta_{13}\sin^2\theta_{12}
\left[
\sin^2\left(\frac{\Delta m^2_{31} L}{4E}\right)
-\sin^2\left(\frac{(\Delta m^2_{31}-\Delta m^2_{21})L}{4E}\right)
\right] \; ,
\end{eqnarray}
where all the mass squared differences and the angles are 
measured at the Earth for antineutrinos. Note that for
antineutrinos, the sign of $V_{cc}$ as well as $V_{e\mu}$
is reversed with respect to the neutrinos.

\begin{figure}
\psfig{file=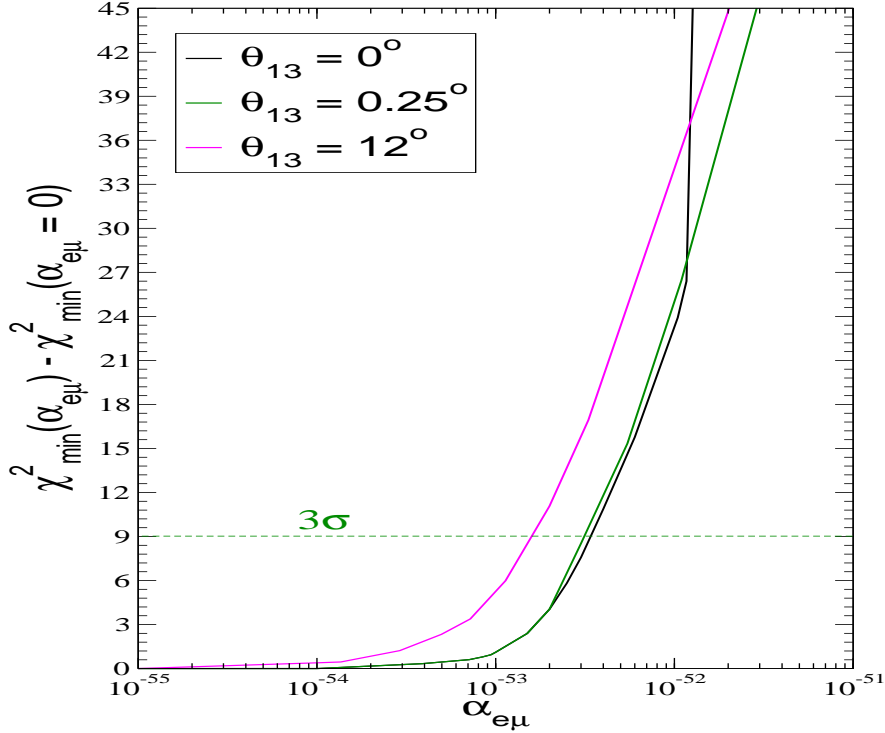,height=10cm,width=12cm}
\caption{  $\Delta \chi^2 \equiv \chi^2(\alpha_{e\mu})-
\chi^2(\alpha_{e\mu}=0)$ values for
different $\theta_{13}$ values,
in the case 
$R_{LR} \ll R_{\rm gal}$.
\label{theta13-dep}}
\end{figure}

In Fig.~\ref{theta13-dep}, we show the 
$\Delta \chi^2$ values as a function of the parameter $\alpha$
for various $\theta_{13}$ values.  
The best fit values for the solar parameters are always observed
to lie in the LMA range with vanishing $\alpha_{e\mu}$ giving
the best fit.
For $\alpha < 10^{-52}$, the value of $\chi^2$ is minimum for
$\theta_{13}=0^\circ$, which is consistent with the 
observation that $\theta_{13}=0^\circ$ also gives the
best fit to the solar and KamLAND data when the LR forces
are not taken into account \cite{lma}.
When $\alpha > 10^{-52}$, a strong energy dependence in
the survival probability is introduced for $\theta < 0.8^\circ$
through $P_H$,
so that the $\chi^2$ values for extremely low $\theta_{13}$
values become large. In this region, the lowest $\chi^2$ is
found to be at values of $\theta_{13}$ that are small, but
still keep $P_H \approx 0$. 
We have shown $\chi^2$ corresponding to such a $\theta_{13}$
in the figure.

\begin{figure}
\centerline{\psfig{figure=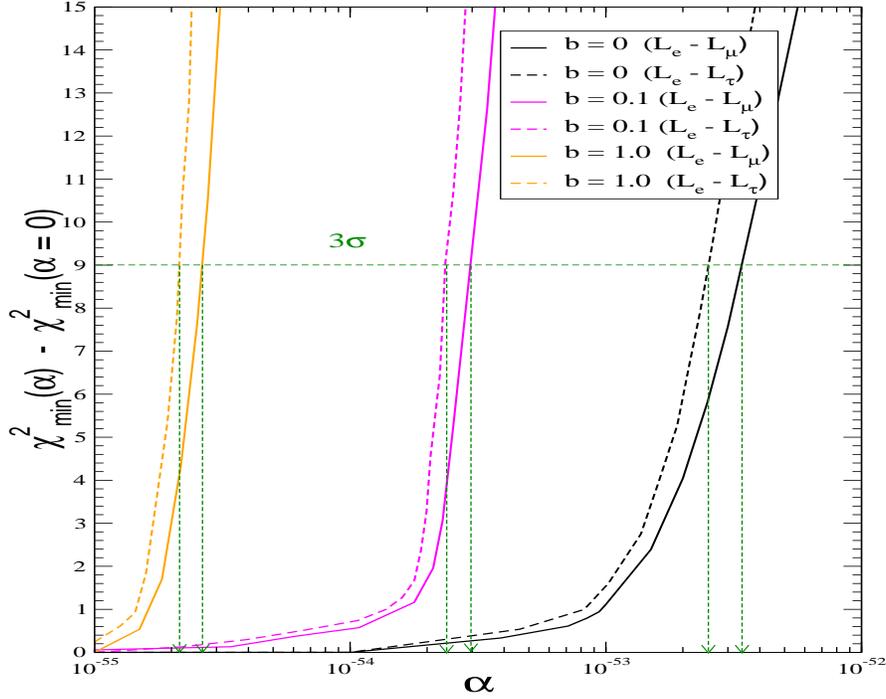,height=10cm,width=12cm,angle=0}}
\caption{  $\Delta \chi^2$ values and limits for the $L_e-L_\mu$ 
as well as $L_e - L_\tau$ symmetry,
with $\theta_{13}=0^\circ$. 
The case $R_{LR} \ll R_{\rm gal}$
is represented by $b=0$ and higher $b$ values
correspond to larger contributions from galactic
electrons (see Sec.~\ref{galaxy}).
\label{chisq-emu}}
\end{figure}

The bounds on $\alpha$ should therefore be, strictly speaking,
$\theta_{13}$-dependent. 
However, the region $\alpha > 10^{-52}$, where the $\theta_{13}$
dependence from $P_H$ starts coming into picture,
is excluded to more than 3$\sigma$ as can be seen 
from Fig.~\ref{theta13-dep}.
Therefore the constraints on $\alpha$ by using
$\theta_{13} = 0^\circ$
are the most conservative ones, 
and we quote the upper bounds on $\alpha$ obtained by
taking $\theta_{13} = 0^\circ$. 
These limits are shown in Fig.~\ref{chisq-emu}:
the $3\sigma$ limit corresponding to the one-parameter fit is
\beq
\alpha_{e\mu} < 3.4 \times 10^{-53} \;.
\eeq
The corresponding limit in the $L_e-L_\tau$ case 
(see Appendix \ref{le-ltau}) is
\beq
\alpha_{e\tau} < 2.5 \times 10^{-53} \;.
\eeq
The bounds are independent of whether the neutrino mass hierarchy is 
normal or inverted.

\section{The long range potential due to galactic electrons}
\label{galaxy}

\begin{figure}
\centerline{\psfig{figure=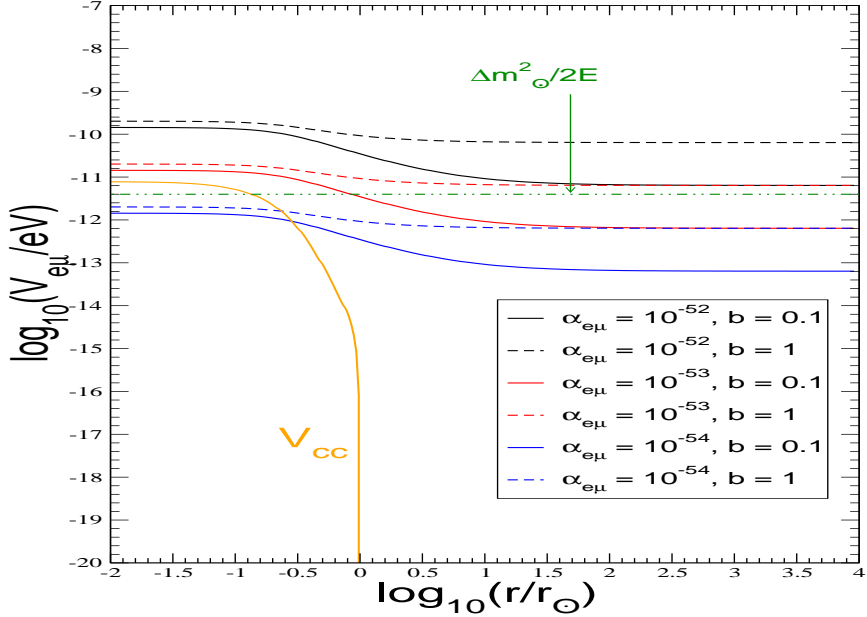,height=9cm,width=12cm,angle=0}}
\caption{The net potential
$V_{e\mu} \equiv V_{e\mu}^\odot + 
V_{e\mu}^{\rm gal}$ for various values of $b$ and $\alpha$.
\label{vemu-net}}
\end{figure}

The collective contribution of all the electrons in the galaxy
to the LR potential in the solar system may be parametrized 
in general as given in eq.~(\ref{b-def}). The net potential
$V_{e\mu}  \equiv V_{e\mu}^\odot + 
V_{e\mu}^{\rm gal}$ is shown in Fig.~\ref{vemu-net} for
various values of $b$ and $\alpha$. 
Clearly, larger the value of $b$ or $\alpha$, larger the
value of $V_{e\mu}$.
Also, note that the value of
$V_{e\mu}$ near the earth is approximately 
the same as $V_{e\mu}^{\rm gal}$, since $V_{e\mu}^\odot$ keeps on
decreasing as one travels towards the Earth, whereas 
 $V_{e\mu}^{\rm gal}$ is a constant 
over the scale of the solar system.

With our understanding of the effects of the LR potential
on neutrino masses, mixings and resonances obtained in 
Sec.~\ref{analytic}, the following observations may be made: \\
(i) For $V_{e\mu}^{\rm gal} \gg \dmsq_\odot /(2E)$,
there is no MSW resonance that is essential for a good fit 
to the solar neutrino data. Therefore, larger values of 
$b$ and $\alpha$ 
($b \alpha \gsim 10^{-53}$) are expected to be
ruled out from the global fit.\\
(ii) For  $V_{e\mu}^{\rm gal} \ll \dmsq_{\rm solar}/(2E)$,
the matter potential $V_{CC}$ dominates inside the Sun, and 
the standard picture of the neutrino flavor conversions
inside the Sun is not affected. Therefore, smaller values of
$b$ and $\alpha$ ($b \alpha \lsim 10^{-55}$) should be allowed.\\
(iii) For the intermediate values of $V_{e\mu}^{\rm gal}$,
the situation depends strongly on whether the potential profile
near the MSW resonance is dominated by $V_{CC}$ or 
$V_{e\mu}^{\rm gal}$. In the former case, the resonance is
adiabatic for $E > 5$ GeV and only partially adiabatic for
lower energies, which gives a good fit to the data. In the
latter case, however, the resonance tends to be adiabatic 
even for low energies, so that the radiochemical data will
disfavor the solution.

The value of $b$ is expected to be in the
range $0.05 < b < 0.5$ (see Sec.~\ref{formalism}).
The $\Delta \chi^2$ values as a function of $\alpha$ for
$b=0 ~(i.e.\ R_{LR} \ll R_{\rm gal})$, $b=0.1$ and $b=1$
are shown in Fig.~\ref{chisq-emu}.
The 3$\sigma$ constraints for $L_e - L_\mu$ are
\beq
\alpha_{e\mu} < 2.9 \times 10^{-54} ~~(b=0.1) \quad , \quad
\alpha_{e\mu} < 2.6 \times 10^{-55} ~~(b=1) 
\label{mulimits}
\eeq
and for $L_e-L_\tau$, they are
\beq
\alpha_{e\tau} < 2.3 \times 10^{-54} ~~ (b=0.1) \quad , \quad
\alpha_{e\tau} < 2.1 \times 10^{-55} ~~ (b=1) \; . 
\label{taulimits}
\eeq
Clearly, the constraints get stronger as $b$ increases.
The most conservative constraints are therefore with
$b=0$, as calculated in Sec.~\ref{numerical}.

\section{Summary and conclusions}
\label{summary}

Flavor-dependent long range leptonic forces,
like those mediated by the $L_e-L_\mu$ or $L_e -L_\tau$ gauge bosons, 
constitute a minimal extension of the 
standard model that preserves its renormalizability. 
The flavor dependent potentials produced by these forces
influence neutrino 
oscillations. The effects of these are quite significant in spite of  
the very strong constraints on the couplings of such forces
from astronomical observations or
E\"otv\"os type laboratory experiments. 
We have performed a 
detailed study of specific effects of these forces in the solar 
neutrino and the KamLAND experiments.

It was found that the new forces change  the standard MSW
picture in a qualitatively different way which ultimately results in a 
strong bound on the couplings of these forces. 
We have developed a detailed
formalism to describe these effects and have used it to obtain 
bounds on the couplings from 
the statistical analysis of the experimental data.
It was shown that the mixing among all three generations 
needs to be taken 
into account because of the fact that 
the $L_e-L_{\mu,\tau}$ gauge bosons couple to two 
out of three flavors at a time. 
The changes which result in the MSW analysis were studied both 
analytically as well as numerically 
in the case $R_{LR} \ll R_{\rm gal}$, when
the galactic electron contribution to the LR potential
may be neglected compared with the solar electron
contribution.

A qualitatively new effect studied in 
detail is the possible 
resonant enhancement of $\theta_{13}$. In the standard MSW picture, a 
non-zero but small  
$\theta_{13}$ can only give sub-leading corrections. 
In contrast, the long range potential 
can resonantly amplify $\theta_{13}$ if $\alpha \gsim 10^{-52}$
and the neutrino mass hierarchy is normal.
The global analysis of the solar data however constrains 
$\alpha< 10^{-52}$. 
As a result, the resonance enhancement of $\theta_{13}$ 
does not take place in
the solar case. But this resonance effect can play an important
role in other environments, e.g. inside a supernova.
See Appendix B for details.

A global $\chi^2$ analysis of all the 
solar neutrino and KamLAND data was 
performed to  constrain the coupling $\alpha$. The solar data alone
are found to be inadequate in constraining $\alpha$: one could 
always
fit these data by appropriate change in $\Delta m_{12}^2$ compared to 
the standard LMA values. This does not remain true when
the KamLAND results are included. A
significant bound on
$\alpha$ is obtained by combining the solar and KamLAND results. 
The conservative $3\sigma$ bounds follow when  $\theta_{13}=0$:
\beq \label{solarbound}
\alpha_{e\mu} < 3.4 \times 10^{-53} \quad , \quad
\alpha_{e\tau} < 2.5 \times 10^{-53} \; .
\eeq
These bounds are stronger by more than one order of magnitude 
than the ones in  eq.~(\ref{atmbound}) 
following from the analysis of the atmospheric neutrino data.

A much stronger bound on $\alpha$, 
namely $\alpha< 6.4\times 10^{-54}$, was quoted in 
\cite{masso} purely from the solar neutrino results. This was not based
on the detailed statistical analysis as presented here, 
but was obtained under 
the assumption that even in the presence of 
$\alpha$ the $\Delta m_{12}^2$
and $\theta_{12}$ should lie within their 95\% C. L. 
range obtained in 
the standard LMA solution. This assumption need not 
{\it a priori} be true.
In fact as discussed here, the solar
neutrino results by themselves cannot be used to 
constrain $\alpha$, so the
detailed analysis as done here is required. 
An analysis similar to ours has
recently been carried out in \cite{garcia} which has 
reported bounds on
couplings of the vector and non-vector long range forces.
The resulting bound on $\alpha$ in the former case 
is similar to ours.
However, they assume one mass scale dominance,
neglecting the mass eigenstate $\nu_3$ altogether.
As shown in this paper, in the presence of the LR potential,
$\nu_3$ affects the solar neutrino survival
probability significantly even when $\theta_{13}$ is 
vanishingly small.
Moreover, the galactic contribution has not been included 
in the analysis of \cite{garcia} even when the range of 
the force is more than our distance
from the galactic center.

When $R_{LR} \gsim R_{\rm gal}$, the collective contribution of 
all the electrons in the galaxy to the LR potential becomes
significant. This gives more stringent constraints on the
value of $\alpha$, which also depend on the distribution of 
baryonic mass within the galaxy. We parametrize our ignorance
about this with a parameter $b$ 
(expected to lie between 0.05 and 1 with conservative estimates) 
and perform global fits
to constrain $\alpha$ for fixed $b$ values. 
We obtain
$\alpha_{e\mu} < 2.9 \times 10^{-54}$ for $b=0.1$ and
$\alpha_{e\mu} < 2.6 \times 10^{-55}$ for $b=1$ in the
$L_e-L_\mu$ case. In the $L_e-L_\tau$ case, one gets
$\alpha_{e\mu} < 2.3 \times 10^{-54}$ for $b=0.1$ and
$\alpha_{e\mu} < 2.1 \times 10^{-55}$ for $b=1$.
Clearly, the constraints become stronger as the galactic
electron contribution, or the range of the 
potential, increases.

The strength of the LR forces increases with the electronic 
content of the source 
and therefore their effects are expected to be much stronger 
for supernova neutrinos.
As discussed in Appendix B, 
the conventional flavor conversions of the supernova neutrinos
changes  significantly in this case 
even for $\alpha\sim 10^{-54}$. In particular, the LR induced 
resonance 
remains adiabatic for very low values of $\theta_{13}$ 
and the Earth matter 
effects may be absent. 
Also, the shock wave effects on the neutrino spectra
may be absent for $t<10$ s, which is when the neutrino flux 
is significant.
On the other hand, the observation of any of these effects may
be used to improve the bound on $\alpha$ at the level of $10^{-54}$,
even when the galactic contribution to the LR forces is small.

While the existence of LR forces may be regarded as a theoretically
allowed speculation at this stage, it is quite remarkable that 
these forces,
if they exist, strongly influence the atmospheric and solar neutrino
oscillations. They would also effect the long baseline experiments which 
can provide additional constraints on $\alpha$.

We have concentrated on bounds on the gauge coupling $\alpha$ 
of the LR forces. 
In principle, the gauge symmetry allows mixing between the
$L_e-L_{\mu,\tau}$ gauge boson $X$ and the ordinary 
hypercharge gauge boson $B$ in their kinetic energy terms. 
This mixing would lead to mixing
between the $X$ boson and the photon and would lead to 
a flavor dependent
infinite range potential even if the $X$ boson has a finite mass. 
The strength of this force will be governed by
an independent mixing parameter $\zeta$ times the electromagnetic 
coupling $\alpha$. 
Based on the present analysis, we expect this quantity to obey
the same constraint as obeyed by $\alpha_{e\mu,\tau}$ 
in case of the infinite range potential.

\section*{Acknowledgments}

AD would like to thank B. Dasgupta and G. Raffelt for 
useful discussions and comments on the manuscript. 
ASJ would like to thank Subhendra Mohanty
for introduction to this subject and for many discussions,
and Tata Institute of Fundamental Research for hospitality.
The work of AB and AD is partly supported through the 
Partner Group project between the Max Planck Institute for
Physics and Tata Institute of Fundamental Research.

\appendix

\section{Constraints on the $L_e-L_\tau$ gauge boson coupling}
\label{le-ltau}

The analysis of $L_e-L_\tau$ gauge bosons can be carried out in an
analogous manner. 
The potential in the flavor basis becomes
\beq \label{tau-v}
V = {\rm Diag}(\vcc + V_{e\tau},0, -V_{e\tau})
\eeq
and the relevant expressions for the mixing angles
in matter (to the leading order in $x$ and $s_{13}$) are:
\barr
\tan 2 \theta_{23m} & \approx & \frac{\sin 2\theta_{23}(1-x
c_{12}^2) }{\cos 2\theta_{23}(1-x c_{12}^2)-y_{e\tau}}~,
\label{23m-etau} \\
\tan 2 \theta_{13m}  & \approx &  \frac{ 2 
(x s_{12} c_{12} S +  s_{13} C)}{C^2+x(c_{12}^2 S^2- s_{12}^2) 
-y_c-y_{e\mu}(1+\cos^2 \theta_{23m})}~,
\label{13m-etau} \\
\tan 2 \theta_{12m} & \approx & \frac{ 2
(x s_{12} c_{12} C - s_{13} S)}{S^2 + x (c_{12}^2 C^2 - s_{12}^2) 
-y_c-y_{e\mu}(1+\sin^2 \theta_{23m})} \; .
\label{12m-etau} 
\earr
The resonance structure is similar to that in the $L_e-L_\mu$
case. The limits on the coupling of such gauge bosons are shown
in Fig.~\ref{chisq-emu}.
The bounds are independent of whether the neutrino mass hierarchy is 
normal or inverted, like in the case of $L_e - L_\mu$.

\section{Effect on a core collapse supernova}
\label{supernova}

The bounds on the LR forces that we obtained from 
the atmospheric, solar and KamLAND experiments are of the
order $\alpha \sim 10^{-53}$, when the galactic contribution
to the LR forces is small.
Although these bounds seem
very stringent, even such a small strength of LR forces
can potentially give rise to significant effects in the
neutrino spectra from a core collapse supernova.
The spectra of $\nu_e$ and $\bar\nu_e$ from the SN
have encoded information about the primary neutrino fluxes
and neutrino mixing parameters \cite{sn-review}, and they
can even show signatures of the passage of the shock wave
through the mantle \cite{shock-effects}.
Note that all the above analyses have been carried out
assuming that the collective flavor conversion effects
caused by the neutrino-neutrino interactions are negligible
compared to the conventional
non-neutrino matter effects on neutrino propagation.
If the collective effects happen to be strong, as claimed in
\cite{collective}, our estimations in this section, as well as 
most of the SN flavor conversion analyses till now need to be
reexamined.

In Fig.~\ref{snfig}, we show a typical profile \cite{profile}
of the MSW potential $V_{cc}$ inside a SN as well as the 
profile of the LR potential $V_{e\mu}$ for two values of $\alpha$
that are allowed with the constraints found
in the conservative scenario $R_{LR} \ll R_{\rm gal}$.
Note that even with $\alpha$ as low as $10^{-54}$, the
LR potential $V_{e\mu}$ exceeds $V_{cc}$ inside the star, 
and hence affects the dynamics of neutrino flavor
conversions. The effects, which may be significant in the
allowed range $10^{-54} < \alpha < 3 \times 10^{-53}$, will be 
as follows:

\begin{figure}
\centerline{\psfig{figure=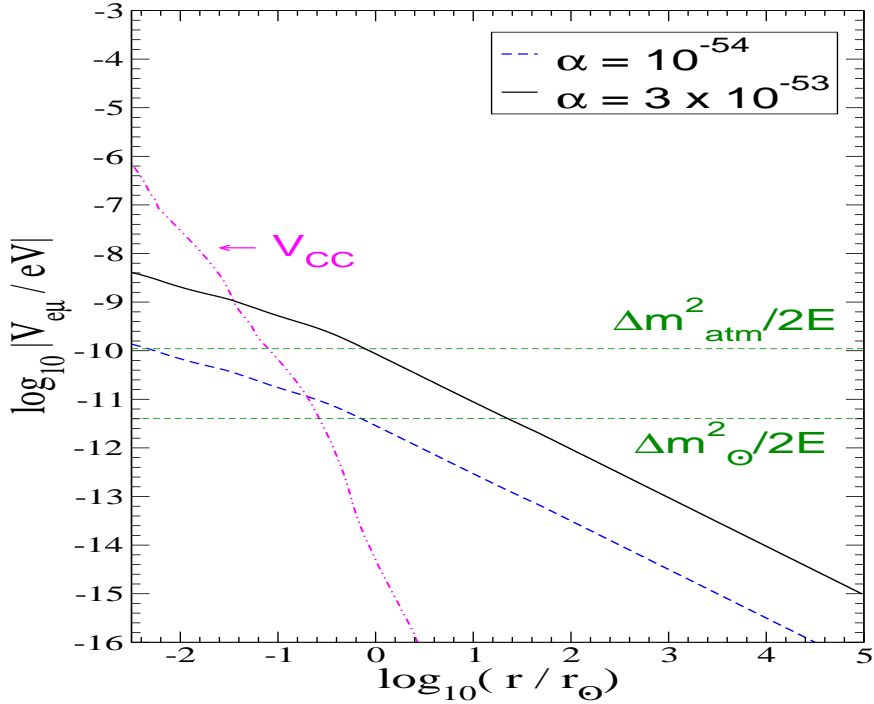,height=10cm,width=12cm,angle=0}}
\caption{The MSW potential and the LR potential for a typical SN
for two different $\alpha$ values that are allowed with the
bounds obtained in this paper for $R_{LR} \ll R_{\rm gal}$. 
The SN model taken \cite{profile}
is a star with a mass of $15 M_\odot$ and primordial metallicity
equal to that of the Sun. 
\label{snfig}}
\end{figure}

\noindent 
(i) The positions of the $H$ and $L$ resonances
\cite{kuo}, corresponding to $\dmsq_\odot$ and $\dmsq_{\rm atm}$
respectively, are shifted away from the center of the star by
a factor of up to one order of magnitude. 

\noindent (ii) If $V_{e\mu}$ dominates over $V_{cc}$ in the resonance
region, the resonance is highly adiabatic, since the 
LR potential is in general smoother than the
MSW potential. Therefore for larger $\alpha$
values, both the $H$ as well as $L$ resonances are
adiabatic for practically all values of $\theta_{13}$. The
SN neutrino spectra then lose the ability to reveal
any information about $\theta_{13}$ in the absence of
any shock wave effects. 
For example, no Earth matter effects \cite{earth-effects}
may be observed.

\noindent
(iii) The shock fronts will reach the resonances at late times, 
$t > 10$ s, when the neutrino flux has reduced a lot. 
As a result, the shock wave effects would be much harder
to observe.

\noindent
(iv) On the other hand, if any effects of non adiabaticity,
e.g. Earth matter effects or shock wave effects, are identified
in the neutrino spectra, the bound on $\alpha$ can be
improved by almost an order of magnitude, to
$\alpha \lsim 10^{-54}$. 
Supernova neutrinos thus form the 
most sensitive probe for the LR forces,
at least when their range is smaller than $R_{\rm gal}$.

\noindent
(v) For $R_{LR} \gsim R_{\rm gal}$, the constraints obtained
from the SN observation are expected to be comparable to 
those found from the solar neutrinos and KamLAND, since
it is the approximate condition $(b ~\alpha~ M_{\rm gal}/R_{\rm gal}) 
\ll \dmsq_{\rm solar} / (2E)$ that determines the allowed 
range of $\alpha$, like in Sec.~\ref{galaxy}.


\end{document}